\documentclass[aps,showpacs,preprintnumbers,nofootinbibt,twocolumn]{revtex4}
\usepackage{amsmath}
\usepackage{eurosym}
\usepackage{amssymb}
\usepackage{graphicx}
\usepackage{color}

\def\be{\begin{equation}}
\def\ee{\end{equation}}
\def\bea{\begin{eqnarray}}
\def\eea{\end{eqnarray}}

\begin{document}

\title{The minimum mass of a spherically symmetric object in $D$-dimensions, and its implications for the mass hierarchy problem}
\author{Piyabut Burikham}
\email{piyabut@gmail.com}
\affiliation{High Energy Physics Theory Group, Department of Physics,
Faculty of Science, Chulalongkorn University, Phyathai Rd., Bangkok 10330, Thailand}
\author{Krai Cheamsawat}
\email{kraicwt@gmail.com}
\affiliation{High Energy Physics Theory Group, Department of Physics,
Faculty of Science, Chulalongkorn University, Phyathai Rd., Bangkok 10330, Thailand}
\author{Tiberiu Harko}
\email{t.harko@ucl.ac.uk}
\affiliation{Department of Mathematics, University College London, Gower Street, London
WC1E 6BT, United Kingdom.}
\author{Matthew J. Lake}
\email{matthewj@nu.ac.th}
\affiliation{The Institute for Fundamental Study, ``The Tah Poe Academia Institute",
\\
Naresuan University, Phitsanulok 65000, Thailand and \\
 Thailand Center of Excellence in Physics, Ministry of Education,
Bangkok 10400, Thailand }

\date{\today }

\begin{abstract}
\vspace{5mm}
The existence of both a minimum mass and a minimum density in nature, in the presence of a positive cosmological
constant, is one of the most intriguing results in classical general relativity. These
results follow rigorously from the Buchdahl inequalities in four dimensional de Sitter space. In this work, we obtain the generalized Buchdahl inequalities in arbitrary space-time dimensions with $\Lambda \neq 0$ and consider both the de Sitter and anti-de Sitter cases. The dependence on $D$, the number of space-time dimensions, of the minimum and maximum masses for stable spherical objects is explicitly obtained. The analysis is then extended to the case of dark energy satisfying an arbitrary linear barotropic equation of state. The Jeans instability of barotropic dark energy is also investigated, for arbitrary $D$, in the framework of a simple Newtonian model with and without viscous dissipation, and we determine the dispersion relation describing the dark energy$-$matter condensation process, along with estimates of the corresponding Jeans mass (and radius). Finally, the quantum mechanical implications of mass limits are investigated, and we show that the existence of a minimum mass scale naturally leads to a model in which dark energy is composed of a `sea' of quantum particles, each with an effective mass proportional to $\Lambda^{1/4}$.


\end{abstract}


\maketitle

\section{Introduction}

The problem of the maximum mass-radius ratio of a stable compact object is one of the most fundamental problems in both general relativity and theoretical astrophysics. In a classic paper, Buchdahl \cite{Buch} obtained the famous result that the ratio of the total mass $M$ and radius $R$ of a high density stable star cannot exceed the value 4/9, $GM/c^2R<4/9$.  The two basic physical assumptions used in the derivation of the upper bound for the mass-radius ratio are that the energy density in the star does not increase outwards and that the pressure is isotropic. On the other hand, by applying the principle of causality and Le Ch{\^a}telier's principle, in \cite{Ruf} it was shown through numerical integration of the general relativistic hydrostatic equilibrium equation (the Tolman-Oppenheimer-Volkoff (TOV) equation) that  the maximum mass of the equilibrium configuration of a dense star cannot exceed $3.2M_{\odot}$, where $M_{\odot} \approx 1.981 \times 10^{33}$ g is the solar mass. This numerical value is presently adopted in the astrophysical literature as indicating the mass limit separating black holes from stable stellar type configurations.

Due to its major astrophysical and theoretical importance, the Buchdahl limit has been extensively investigated. The effects of the presence of a cosmological constant on the stellar mass-radius ratio were considered in \cite{Har1}, while limits on $M/R$ for charged spheres were derived in \cite{Har2}. In \cite{Andr1}, it was argued that some of the assumptions used to derive the Buchdahl inequality were very restrictive. For example,  neither of them hold for a simple soap bubble. By relaxing these assumptions and considering any static solution of the spherically symmetric Einstein equations for which the energy density $\rho \geq 0$ and the radial and tangential pressures, $p\geq 0$ and $p_T$, satisfy the condition $p+2p_T \leq \Omega \rho c^2 $, $\Omega >0$, one can obtain the relation $\sup _{r>0}\left[2Gm(r)/c^2r\right]\leq \left[\left(1+2\Omega\right)^2-1\right]/\left(1+2\Omega \right)^2$ \cite{Andr1}. These bounds were generalized to the case of charged compact general relativistic objects in \cite{Andr2}. Bounds on $M/R$ for static objects with a positive cosmological constant $\Lambda >0$ were obtained in \cite{Andr3}, where it was shown that the relation $GM/c^2R\leq 2/9-\Lambda R^2/3+(2/9) \sqrt{1+3\Lambda R^2}$ holds if the energy conditions listed above are satisfied. Buchdahl type inequalities, expressed in terms of the mean fluid density of the sphere, in space-times with arbitrary $D$ and $\Lambda \neq 0$ were also derived in \cite{Ponce, Zarro,Wright1}, while the case of stable stars in five dimensional Gauss-Bonnet gravity was considered in \cite{Wright2}. In \cite{Ponce}  it was shown that in $D$ dimensions the Buchdahl inequality for the maximum mass-radius ratio can be  formulated as $GM/R^{D-3}\leq 2(D-2)/(D-1)^2$. The standard assumptions used in deriving the Buchdahl inequality were relaxed in \cite{Wright1}, where various matter property depending bounds were obtained.

In \cite{Boehmer:2005sm} it was shown that, in the framework of classical general relativity, the presence of a positive cosmological constant implies the existence of a minimum mass and a minimum density in nature. These results follow rigorously from the generalized Buchdahl inequalities for $D=4$, with $\Lambda>0$. In this scenario, the mass and mean density of a stable compact object must satisfy the conditions $2GM/c^2R \geq \Lambda R^2/6$ and $\bar{\rho} \geq \Lambda c^2/16\pi G$, respectively.  These bounds were extended to the case of anisotropic compact objects in \cite{Boehm2}, while bounds for the minimum masses of charged particles were derived in \cite{Boehmer:2007gq}. The physical implications of the existence of a minimum mass (in $D=4$) were investigated in \cite{Boehm3}. An especially important result is that the ratio $l_{Pl}^4/\Lambda$, where $l_{Pl}$ is the reduced Planck length and $\Lambda$ is the four-dimensional cosmological constant, is numerically of the same order of magnitude as $r_e^6$, where $r_e \approx 2.818 \times 10^{-13}$ cm denotes the classical electron radius. This suggests the identification of $\Lambda$ in terms of fundamental physical constants as
\begin{equation}\label{const}
\Lambda = \frac{l_{Pl}^4}{r_e^6}=\frac{\hbar ^2G^2m_e^6c^6}{e^{12}}\approx 1.4\times 10^{-56}\;{\rm cm}^{-2},
\end{equation}
where  $m_e$ and $e$ are the electron mass and charge, respectively  \cite{Boehm3}. Remarkably, the same formula was obtained using a set of axioms for $\Lambda$ based on a close analogy with the Khinchin axioms in information theory \cite{Khinchin}. In this method, the dependency of the information measure on probabilities of events was formally replaced by the dependency of the cosmological constant on the fundamental constants of nature \cite{Beck:2008rd}.

In Section \ref{quant}, we also consider a new interpretation of the bound on $l_{Pl}^4/\Lambda$, in terms of the Chandrasekhar mass for a condensate of quantum dark energy particles. An intriguing possibility is that dark energy, obeying an equation of state of the form $\rho _{DE}c^2+p_{DE}=0$, may condense to form stable self-gravitating objects. This scenario was originally considered in \cite{ds1}-\cite{ds7}.  Dark energy stars or, in a wider sense, objects with negative pressure in their interiors, are interesting alternatives to the standard black hole paradigm. In one implementation of this idea, hypothetical compact general relativistic objects called gravastars ({\it gra}vitational {\it va}cuum stars) have been proposed as an alternative explanation for the astrophysical characteristics usually associated with black holes \cite{grav1}-\cite{grav8}. The basic physical idea of this scenario is that the  quantum vacuum undergoes a phase transition at the moment the event horizon is formed.
Therefore, the structure of a gravastar consists of an interior de Sitter condensate obeying the dark energy equation of state, $\rho c^2=-p$. This interior is  matched to an exterior consisting of a shell of finite thickness described by the equation of state $\rho c^2=p$, and the shell is then matched at its vacuum boundary to an exterior Schwarzschild solution.

In this paper, we consider the Buchdahl limit, and the resulting minimum mass and density, for static spherically symmetric compact objects in an arbitrary $D$-dimensional geometry. From the Einstein field equations in arbitrary dimensions with $\Lambda \neq 0$ and the hydrostatic equilibrium equation, the generalizations of the Buchdahl limit for arbitrary $D$, and of the minimum mass allowed for any classical elementary particle, are obtained. In the particular case of $D=4$, these limits reduce to the corresponding expressions given in \cite{Buch} and \cite{Boehmer:2005sm}, respectively. On the other hand, cosmological observations such as those of high redshift supernovae or the Cosmic Microwave Background (CMB) data from the Planck mission \cite{Planckresults,Bet} suggest that the dark anergy equation of state is linear, with the state parameter lying in the range $ -1 < w =p_{DE}/\rho _{DE} < -1/3$, where $p_{DE}$ and $\rho _{DE}$ are the thermodynamic pressure and the dark energy density, respectively \cite{data}. Therefore, the possibility that dark energy is not {\it exactly} a cosmological constant cannot be rejected a priori. Taking into account this possibility, we obtain the Buchdahl and minimum mass limits in arbitrary space-time dimensions for dark energy obeying a linear barotropic equation of state. The conditions for the collapse of a star embedded in the dark energy fluid are also derived in both the de Sitter and anti-de Sitter cases.

In addition, an interesting physical possibility is that dark energy may undergo a phase transition leading to a condensation process, which could be either gravitational or of Bose-Einstein type. In this scenario, the condensation of dark energy `particles' could produce compact super-massive objects. We study the condensation process using the classical method of Jeans instability \cite{BT}, generalized to arbitrary space-time dimensions, and by taking into account the possibility of the presence of viscous dissipative effects in the dark energy fluid. The role of the dissipative processes in the occurrence of the Jeans instability was studied in \cite{visc}. As a first step we derive the mass of the dark energy condensate in the framework of the Newtonian dissipationless approximation by assuming that the dark energy fluid condenses, or transforms via a phase transition, into an ideal non-relativistic fluid. In four dimensions, the corresponding Jeans mass is proportional to $\Lambda ^{-1/2}$ and its numerical value is close to the observationally estimated total mass of the Universe. We also show that the Jeans mass can be represented in a form similar to the Chandrasekhar mass, that depends only on the fundamental constants. The effect of the bulk viscosity of the dark energy fluid on the condensation process is also investigated.

Although a self-consistent and credible theory of quantum gravity has not been found, we use general arguments to investigate the possible quantum mechanical implications of the existence of a minimum mass and minimum density. Firstly, we show that, in four-dimensions, the Jeans mass of the dark energy  $M_J\propto \left(c^2/G\right)\Lambda^{-1/2}$, having a numerical value of the order of the mass of the Universe,  can be obtained from thermodynamic considerations related to the physics of Schwarzschild-de Sitter black holes. We then associate the minimum mass with a temperature which, intriguingly, is very close to the present day temperature of the CMB radiation. Moreover, we show that the existence of a minimum mass bound leads naturally to a hierarchical model in which quantum `particles' above a certain size effectively decohere via interaction with the cosmological constant (or dark energy fluid) and in which dark energy itself is composed of quantum particles with an effective mass proportional to $\Lambda^{1/4}$.

This paper is organized as follows. In Section~\ref{masss}, the $D$-dimensional Einstein field equations with a nonzero cosmological constant are introduced and the expressions for the Buchdahl limit and the classical minimum mass are obtained. The more general case of a $D$-dimensional sphere of matter embedded in a dark energy fluid obeying an arbitrary barotropic equation of state is considered in Section~\ref{massdes}, and the corresponding limiting masses are derived. The Jeans instability of the dark energy fluid is discussed in Section~\ref{Jeans}, within the framework of a simple Newtonian model. Finally, the quantum mechanical implications of the existence of a classical minimal mass and density are investigated in Section~\ref{quant} and a brief discussion of our main results is presented in Section~\ref{concl}.

\section{Mass limits for spherically symmetric stars  in $D$ dimensions in
the presence of a cosmological constant}\label{masss}

In this Section, we first introduce the Einstein field equations in arbitrary space-time dimensions with $\Lambda \neq 0$ and derive the hydrostatic equilibrium equation (the Tolman-Oppenheimer-Volkoff (TOV) equation) for spherically symmetric objects. We then derive the $D$-dimensional generalizations of the Buchdahl limit and of the minimum mass in general relativity. In estimating critical length and mass scales throughout this paper we use the approximate values $c \approx 2.998 \times 10^{10} \ {\rm cm}{\rm s}^{-1}$, $G \approx 6.674 \times 10^{-8} \;\ {\rm cm}^{3}\;{\rm g}^{-1}{\rm s}^{-2}$, $h \approx 2\pi \times 1.055 \times 10^{-27} \ {\rm erg}\;{\rm s}$ and $\Lambda \approx 3 \times 10^{-56} \ {\rm cm}^{-2}$ for the fundamental constants.

\subsection{Tolman-Oppenheimer-Volkoff equation in $D$ - dimensional space-time}

In arbitrary $D$ space-time dimensions, the interior metric inside a
spherically symmetric fluid sphere takes the form \cite{D}
\begin{equation}  \label{metr}
ds^2 = e^{\nu(r)}d(ct)^2 - e^{\lambda(r)}dr^2 - r^2d\Omega^2_{D-2},
\end{equation}
where $d\Omega^2_{D-2}$ is the line element on the unit $S_{D-2}$ sphere,
\begin{equation}
d\Omega_{D-2}^2 = d\theta_1^2+\sin^2{\theta_1}d\theta_2^2+\ldots + \sin^2{%
\theta_1}\cdots\sin^2{\theta_{D-3}}d\phi^2,
\end{equation}
with the coordinate domains $0 \leq r < \infty$, $0 \leq \theta_i \leq \pi\;(i = 1,\ldots,
D-3) $ and $0 \leq \phi < 2\pi$.

The Einstein field equations in $D$ space-time dimensions, in the
presence of a $D$-dimensional cosmological constant $\Lambda _D$, are
\begin{equation}
R^{\mu}_{\phantom{1}\nu}-\frac{1}{2}\delta^{\mu}_{\phantom{1}\nu}R-\Lambda
_D\delta^{\mu}_{\phantom{1}\nu} = \frac{8\pi G_D}{c^4}T^{\mu}_{\phantom{1}\nu}, \label{efeqn}
\end{equation}
where the $D$-dimensional Newton's constant $G_D$ is implicitly
defined so that the field equations takes the same form in any number of dimensions. The energy momentum tensor inside the fluid sphere is
given by
\begin{equation}
T^{\mu}_{\phantom{1}\nu} = (\rho c^2+P)u^{\mu}u_{\nu}-P\delta^{\mu}_{\phantom{1}%
\nu},
\end{equation}
where $u^{\mu}$ is the $D$-velocity, $u^{\mu} = \delta^{\mu}_{0}e^{-\nu(r)/2}$, $%
\rho$ is the energy density and $P$ is the pressure.

For the metric given by Eq.~(\ref{metr}), the gravitational field equations
become \cite{D, Burikham:2012kn}
\begin{equation}  \label{00}
\frac{(D-2)\lambda^{\prime }e^{-\lambda}}{2r}-\frac{%
(D-2)(D-3)(e^{-\lambda}-1)}{2r^2} = \frac{8\pi G_D}{c^2}\rho + \Lambda _D,
\end{equation}
\be \label{rr}
\frac{(D-2)\nu^{\prime }e^{-\lambda}}{2r}+\frac{%
(D-2)(D-3)(e^{-\lambda}-1)}{2r^2} = \frac{8\pi G_D}{c^4}P - \Lambda _D,
\ee
\bea
&& e^{-\lambda}\left[ \frac{\nu^{\prime \prime }}{2}+\frac{%
\nu^{\prime 2}}{4}-\frac{\nu^{\prime }\lambda^{\prime }}{4}+\frac{%
(D-2)(\nu^{\prime }-\lambda^{\prime })}{4r} \right]+ \nonumber\\
&&\frac{%
(D-3)(D-4)(e^{-\lambda}-1)}{2r^2} = \frac{8\pi G_D}{c^4}P-\Lambda _D,
\eea
where a prime denotes the derivative with respect to the radial coordinate $r$. 
From the continuity equation, $\nabla_{\mu}T^{\mu}_{\phantom{1}\nu} = 0$, it follows that
\begin{equation}
\nu^{\prime }= -\frac{2P^{\prime }}{\rho c^2+P} .  \label{conservation}
\end{equation}
Equation (\ref{00}) can be immediately integrated, giving
\bea
e^{-\lambda(r)} &=& 1-\frac{8\pi G_D}{c^2} \frac{2}{D-2}\frac{1}{r^{D-3}}\int_0^r
\rho(r^{\prime })r^{\prime D-2}dr^{\prime }- \nonumber\\
&&\frac{2\Lambda _Dr^2 }{%
(D-1)(D-2)}.
\eea
The `accumulated mass at radius $r$' (i.e. the mass inside the radius $r$) is
defined in the $D-1$ dimensional spatial volume, in a way consistent with the
four-dimensional, case as
\begin{equation}  \label{meqn}
M(r) = \Omega_{D-2} \int_0^r\rho(r^{\prime })r^{\prime D-2}dr^{\prime },
\end{equation}
where $\Omega_{D-2}$ is the surface area of the unit $S_{D-2}$ sphere,
\begin{equation}
\Omega_{D-2} = \frac{2\pi^{\frac{D-1}{2}}}{\Gamma\left(\frac{D-1}{2}\right)},
\end{equation}
and $\Gamma (x)$ is the Gamma function. For $D=4$, $\Gamma (3/2)=\sqrt{\pi}%
/2 $, and we obtain $\Omega _2=4\pi$.

Therefore
\begin{equation}
e^{-\lambda(r)} = 1-\frac{16\pi G_DM(r)}{(D-2)c^2\Omega_{D-2}r^{D-3}} - \frac{%
2\Lambda _Dr^2 }{(D-1)(D-2)}.  \label{grr}
\end{equation}

By using Eqs.~(\ref{conservation}) and (\ref{grr}) in Eq.~(\ref{rr}), we
obtain the TOV equation in arbitrary space-time dimensions, in the
presence of a cosmological constant, as
\begin{widetext}
\begin{equation}
\frac{dP}{dr}=-\frac{(\rho c^2 +P)\left[ \left( \frac{8\pi G_{D}}{c^4}P-\frac{2\Lambda _{D}}{%
D-1}\right) r^{D-1}+(D-3)\frac{8\pi G_{D}}{c^2\Omega _{D-2}}M(r)\right] }{%
\left( D-2\right) r^{D-2}\left[ 1-\frac{16\pi G_{D}}{(D-2)c^2\Omega
_{D-2}r^{D-3}}M(r)-\frac{2\Lambda _{D}r^{2}}{(D-1)(D-2)}\right] }.
\label{TOV}
\end{equation}%
\end{widetext}
%

\subsection{Upper and lower bounds for the mass-radius ratio in $D$
space-time dimensions}\label{massu}

The bounds on the mass-radius ratio in $D$ dimensions can be derived solely by
imposing some realistic restrictions on the behavior of the energy density $%
\rho $. The mass-radius ratio is independent of the equation of
state that relates $\rho (r)$ and $P(r)$ and, using Eqs.~(\ref{conservation})
and (\ref{TOV}), the thermodynamic pressure $P(r)$ can always be eliminated
from the field equations. It is convenient to first rewrite the equations in terms of the
generalized Buchdahl variables $\left( x,w,\zeta ,y\right) $, defined as
\begin{eqnarray*}
x &=&r^{2} , w(r)=\frac{8\pi G_{D}M(r)}{(D-2)c^2\Omega _{D-2}r^{D-1}}%
,\zeta =e^{\nu /2}, \nonumber\\
y^{2} &=&e^{-\lambda (r)}=1-2w(r)r^{2}-\frac{2\Lambda _{D}r^{2}}{(D-1)(D-2)}.\nonumber\\
\end{eqnarray*}%
In terms of the new variables, Eqs.~(\ref{conservation}) and (\ref{TOV})
can be written as
\begin{equation}
\frac{1}{\zeta }\frac{d\zeta }{dx}=-\frac{1}{\rho c^2 +P}\frac{dP}{dx}%
\rightarrow \frac{d}{dx}(\zeta P)=-\rho c^2 \frac{d\zeta }{dx},
\label{conservation2}
\end{equation}%
and
\bea \label{TOV2}
\frac{dP}{dx}&=&-\frac{(\rho c^2 +P)}{y^{2}}\Bigg[ \frac{1}{(D-2)}\left( \frac{4\pi G_{D}}{c^4}P-\frac{\Lambda _{D}}{(D-1)}\right) +\nonumber\\
&&\frac{(D-3)}{2}w\Bigg],
\eea
respectively. After eliminating $P(r)$, we obtain
\begin{equation}
\frac{d}{dx}\left( y\frac{d\zeta }{dx}\right) -\frac{D-3}{2}\frac{\zeta }{y}%
\frac{dw}{dx}=0.  \label{buchdahl}
\end{equation}%
Note that, in the case of constant density (or for $D=3$), the second term in Eq.~(\ref%
{buchdahl}) vanishes. Throughout the rest of this analysis, we impose the following physical conditions on the stellar structure. We require
that both the local density $\rho$ and the mean density
\begin{equation}
\bar{\rho}=\frac{(D-1)M(r)}{\Omega _{D-2}r^{D-1}},
\end{equation}%
do not increase as $r$ increases. That is, we require $\rho $ and $\bar{\rho}$ to be
monotonically decreasing functions of $r$ inside the fluid matter
distribution. We also assume that this condition is independent of the
equation of state of the matter inside the star. The density monotonicity
condition implies that
\begin{equation}
\frac{d}{dr}\left( \frac{M(r)}{r^{D-1}}\right) <0\quad \Rightarrow \quad
\frac{dw}{dx}<0.
\end{equation}%
Using Eq.~(\ref{buchdahl}) and introducing a new independent variable $\ell $, defined as
\bea
\ell (r)&=&\int_{0}^{r}r^{\prime }\Bigg[ 1-\frac{16\pi G_{D}M(r)}{(D-2)c^2\Omega
_{D-2}r^{D-3}}-\nonumber\\
&&\frac{2\Lambda _{D}r^{2}}{(D-1)(D-2)}\Bigg] ^{-\frac{1}{2}%
}dr^{\prime },
\eea
with $\ell (0)=0$, we obtain the condition for
the stability of a general relativistic fluid sphere in $D$ space-time
dimensions in the following form,
\begin{equation}
\frac{d^{2}\zeta }{d\ell ^{2}}<0.
\end{equation}%
This condition must hold for all $r\in \lbrack 0,R]$, where $R$ marks the spatial boundary of the fluid distribution. By using the mean
value theorem, we obtain the following inequality for $d\zeta/dl$
\begin{equation}
\frac{d\zeta }{d\ell }\leq \frac{\zeta (\ell )-\zeta (0)}{\ell }.
\end{equation}%
Since $\zeta (0)>0$, it follows that
\begin{equation}
\frac{1}{\zeta }\frac{d\zeta }{d\ell }\leq \frac{1}{\ell }.  \label{ineq}
\end{equation}%
Next, we introduce the parameter $\alpha (r)$, defined as
\begin{equation}
\alpha (r)=1+\frac{2\Lambda _{D}}{(D-1)(D-2)}\frac{(D-2)c^2\Omega _{D-2}r^{D-1}%
}{8\pi G_{D}M(r)},
\end{equation}%
so that
\begin{equation}
y^{2}=1-\frac{16\pi G_{D}M(r)\alpha (r)}{(D-2)c^2\Omega _{D-2}r^{D-3}}.
\end{equation}%
From the condition that $\frac{d}{dr}\left( \frac{M(r)}{r^{D-1}}\right) <0$,
we can conclude that, for $r^{\prime }<r$,
\begin{equation}
\frac{M(r^{\prime })}{r^{\prime D-1}}\geq \frac{M(r)}{r^{D-1}}.
\end{equation}%
We can also assume that
\begin{equation}
\alpha (r^{\prime })\frac{M(r^{\prime })}{r^{\prime D-3}}\geq \alpha (r)%
\frac{M(r)}{r^{D-3}}\left( \frac{r^{\prime }}{r}\right) ^{2},
\end{equation}%
or, equivalently,
\bea
&&\left[ 1-\frac{16\pi G_{D}M(r^{\prime })\alpha (r^{\prime })}{(D-2)c^2\Omega
_{D-2}r^{\prime D-3}}\right] ^{-\frac{1}{2}}\geq \nonumber\\
&&\Bigg[ 1-
\frac{16\pi
G_{D}M(r)\alpha (r)}{(D-2)c^2\Omega _{D-2}r^{D-1}}r^{\prime 2}\Bigg] ^{-\frac{1%
}{2}}.
\eea
Hence, the right-hand side of the inequality (\ref{ineq}) is bounded from above, such that
\begin{widetext}
\begin{eqnarray}
\hspace{-0.3cm}\left\{ \int_{0}^{r}r^{\prime }\left[ 1-\frac{16\pi G_{D}M(r^{\prime
})\alpha (r^{\prime })}{(D-2)c^2\Omega _{D-2}r^{\prime D-3}}\right] ^{-\frac{1}{%
2}}dr^{\prime }\right\} ^{-1} &\leq &\frac{16\pi G_{D}M(r)\alpha (r)}{%
(D-2)c^2\Omega _{D-2}r^{D-1}}
\left[ 1-\sqrt{1-\frac{16\pi G_{D}M(r)\alpha (r)}{(D-2)c^2\Omega _{D-2}r^{D-3}%
}}\right] ^{-1}.
\end{eqnarray}%
\end{widetext}
Substituting from Eqs.~(\ref{conservation2}) and (\ref{TOV2}) in Eq. (\ref{ineq}), we finally obtain
\bea\label{ineq2}
&&\frac{1}{D-2}\left( \frac{8\pi G_{D}}{c^4}P-\frac{2\Lambda _D}{D-1}\right) r^{2}+\nonumber\\
&&(D-3)%
\frac{8\pi G_{D}M(r)}{(D-2)c^2\Omega _{D-2}r^{D-3}}\leq y(1+y).
\eea
The Buchdahl inequality (\ref{ineq2}) is valid for all $r \in [0,R]$ inside the star.
This allows us to determine the upper bound on the mass-radius ratio, which is a
natural consequence of the theory of general relativity (there is no similar
bound in Newtonian gravity), and to determine the more subtle lower bound on the
mass-radius ratio when a nonzero cosmological constant is included in the analysis.

To obtain
these bounds, we evaluate the inequality (\ref{ineq2}) at the surface of the
star $r=R$, where $P(R)=0$ and $M(R)=M$, with $M$ denoting the total mass. This gives
\begin{widetext}
\begin{eqnarray}  \label{ineqf}
\hspace{-0.3cm}(D-3)\frac{8\pi G_DM}{(D-2)c^2\Omega_{D-2}R^{D-3}} &\leq &\sqrt{1-\frac{16\pi
G_DM}{(D-2)c^2\Omega_{D-2}R^{D-3}}-\frac{2\Lambda _DR^2}{(D-1)(D-2)}}+
1-\frac{16\pi G_DM}{(D-2)c^2\Omega_{D-2}R^{D-3}}.
\end{eqnarray}
\end{widetext}

For convenience, we redefine the variables in the above inequality as
\begin{equation} \label{ub}
u := \frac{8\pi G_DM}{(D-2)c^2\Omega_{D-2}R^{D-3}}, \;b := \frac{2\Lambda _DR^2%
}{(D-1)(D-2)}.
\end{equation}
Hence the inequality (\ref{ineqf}) becomes
\begin{equation} \label{ineqf*}
(D-3)u \leq \sqrt{1-2u-b} + 1 -2u.
\end{equation}
By squaring both sides of Eq. (\ref{ineqf*}) and simplifying, we obtain
\begin{equation}
u^2 - 2\frac{(D-2)}{(D-1)^2}u + \frac{b}{(D-1)^2} \leq 0,
\end{equation}
or, equivalently,
\begin{equation}  \label{ineq3}
(u-u_1)(u-u_2) \leq 0 ,
\end{equation}
where we have denoted
\begin{eqnarray}
u_1 &=& \frac{D-2}{(D-1)^2}\left[ 1+ \sqrt{1-\frac{2(D-1)}{(D-2)^3}\Lambda
_D R^2} \right], \\
u_2 &=& \frac{D-2}{(D-1)^2}\left[ 1- \sqrt{1-\frac{2(D-1)}{(D-2)^3}\Lambda
_D R^2} \right].
\end{eqnarray}
To check the sign of each factor in the inequality (\ref{ineq3}), we set $%
\Lambda _D= 0$, giving
\begin{equation}
(u-u_1)u \leq 0
\end{equation}
Since $u$ is always positive, the condition $u \leq u_1$ implies that
\begin{eqnarray}
\frac{8\pi G_DM}{(D-2)c^2\Omega_{D-2}R^{D-3}} &\leq& \frac{2(D-2)}{(D-1)^2}, \\
\frac{4\pi G_DM}{c^2\Omega_{D-2}R^{D-3}} &\leq& \left( \frac{D-2}{D-1}
\right)^2.
\end{eqnarray}
When $D=4$, we reproduce the known result represented by the standard
Buchdahl inequality \cite{Buch}.
\begin{equation} \label{Buch_1}
\frac{G_4M}{c^2R} \leq \frac{4}{9}.
\end{equation}
Generically, for nonzero $\Lambda _D$, 
the reality of $u_1$ and $u_2$ requires
\begin{equation} \label{reality_cond}
R^2 \leq \frac{1}{2}\frac{(D-2)^3}{(D-1)}\Lambda_D^{-1},
\end{equation}
and we can conclude that $u$ satisfies
the following inequality
\begin{equation}
u-u_1 \leq 0 \quad , \quad u-u_2 \geq 0.
\end{equation}
Therefore, when a nonzero cosmological constant is taken into account,
there exist both lower and upper bounds on the mass-radius ratio for
stellar type objects in arbitrary $D$ dimensional space-times
\begin{widetext}
\begin{equation}  \label{masb}
\left(\frac{D-2}{D-1}\right)^2 \left[ 1- \sqrt{1-\frac{2(D-1)}{(D-2)^3}%
\Lambda _DR^2} \right] \leq \frac{8\pi G_DM}{c^2\Omega_{D-2}R^{D-3}} \leq \left(%
\frac{D-2}{D-1}\right)^2 \left[ 1+ \sqrt{1-\frac{2(D-1)}{(D-2)^3}\Lambda
_DR^2} \right].
\end{equation}
\end{widetext}
Equation~(\ref{masb}) gives a nontrivial~(positive) lower bound only when $0 < \Lambda _D < (1/2)(D-2)^{3}(D-1)^{-1}R^{-2}$.
On the contrary, the upper bound always exists, regardless of the sign of $%
\Lambda _D$.

\section{Mass limits for spherically symmetric systems in the presence of dark
energy}\label{massdes}

In this Section, we generalize the analysis in Section \ref{masss} to the
situation where the `star' (i.e. spherically symmetric object) is embedded in a space-time filled with dark
energy, characterized by a thermodynamic energy density $\rho _{DE}$ and a
thermodynamic pressure $p_{DE}$, respectively, obeying a generic equation of state $P_{%
\mathrm{DE}}=w\rho _{\mathrm{DE}}c^2$, where $w=\mathrm{constant}$. The
Einstein equations then become
\bea\label{00de}
&&\frac{(D-2)\lambda ^{\prime }e^{-\lambda }}{2r}-\frac{%
(D-2)(D-3)(e^{-\lambda }-1)}{2r^{2}}=\nonumber\\
&&\frac{8\pi G_{D}}{c^2}(\rho +\rho _{\mathrm{DE}}),
\eea
\bea\label{rrde}
&&\frac{(D-2)\nu ^{\prime }e^{-\lambda }}{2r}+\frac{%
(D-2)(D-3)(e^{-\lambda }-1)}{2r^{2}}=\nonumber\\
&&\frac{8\pi G_{D}}{c^4}(P+w\rho_{\mathrm{DE}}c^2),
 \eea
 \bea
&&e^{-\lambda }\left[ \frac{\nu ^{\prime \prime }}{2}+\frac{\nu
^{\prime 2}}{4}-\frac{\nu ^{\prime }\lambda ^{\prime }}{4}+\frac{(D-2)(\nu
^{\prime }-\lambda ^{\prime })}{4r}\right] +\nonumber\\
&&\frac{(D-3)(D-4)(e^{-\lambda }-1)%
}{2r^{2}}=\frac{8\pi G_{D}}{c^4}\left(P+w\rho _{\mathrm{DE}}c^2\right).\nonumber\\
\eea

Equation~(\ref{00de}) can be integrated again to obtain
\begin{equation}
e^{-\lambda (r)}=1-\frac{16\pi G_{D}\left[ M(r)+M_{\mathrm{DE}}(r)\right] }{%
(D-2)c^2\Omega _{D-2}r^{D-3}},  \label{grrde}
\end{equation}%
where the ordinary `matter mass' $M(r)$ is given by Eq.~(%
\ref{meqn}) and the dark energy mass is given by the same relation, with $%
\rho _{\mathrm{DE}}$ replacing $\rho $. Note that the total physical
accumulated mass of the star is thus $M_{tot}(r)=M(r)+M_{\mathrm{DE}}(r)$.
By using the Einstein equations and the conservation of the energy-momentum tensor, the
generalized TOV equation can be obtained as
\begin{widetext}
\begin{equation}
\frac{d(P+w\rho _{\mathrm{DE}}c^2)}{dr}=-\frac{\left[ \rho c^2 +P+(1+w)\rho _{%
\mathrm{DE}}c^2\right] \left[ \frac{8\pi G_{D}}{c^4}\left( P+w\rho _{\mathrm{DE}}c^2\right)
r^{D-1}+\frac{8\pi G_{D}(D-3)}{c^2\Omega _{D-2}}M_{tot}(r)\right] }{\left(
D-2\right) r^{D-2}\left[ 1-\frac{16\pi G_{D}}{(D-2)c^2\Omega _{D-2}r^{D-3}}%
M_{tot}(r)\right] }.  \label{TOVde}
\end{equation}%
\end{widetext}

We now assume that the dark energy profile is constant throughout the star, which
is equivalent to assuming the ultrastiff condition for the dark energy fluid. For a
constant dark energy density profile, the dark energy mass content of the
star becomes
\begin{equation}
M_{\mathrm{DE}}(r)=\Omega _{D-2}\rho _{\mathrm{DE}}\frac{r^{D-1}}{D-1}.
\end{equation}%
Hence the TOV equation can be rewritten as
\begin{widetext}
\begin{equation}
\frac{dP_{eff}}{dr}=-\frac{(\rho c^2 +P_{eff})\left[ \left( \frac{8\pi G_{D}}{c^4}P_{eff}-%
\frac{2\Lambda_D }{D-1}\right) r^{D-1}+(D-3)\frac{8\pi G_{D}M(r)}{c^2\Omega _{D-2}%
}\right] }{\left( D-2\right) r^{D-2}\left[ 1-\frac{16\pi G_{D}M(r)}{%
(D-2)c^2\Omega _{D-2}r^{D-3}}-\frac{2\Lambda_D r^{2}}{(D-1)(D-2)}\right] },
\label{TOVdec}
\end{equation}%
\end{widetext}
where we have denoted $\Lambda _{D}\equiv 8\pi G_{D}\rho _{\mathrm{DE}}/c^2$,
and the effective pressure is $P_{eff}\equiv P+(1+w)\rho _{\mathrm{DE}}c^2$.
Note that Eq. (\ref{TOVdec})  is simply Eq.~(\ref{TOV}) for $P_{eff}$.

As a result, the same analysis as performed in Section \ref{massu} can be
repeated to obtain
\begin{widetext}
\begin{equation} \label{ineqdef}
\frac{\Lambda _{D}R^{2}}{D-2}(1+w)+\frac{(D-1)8\pi G_{D}M}{(D-2)c^2\Omega
_{D-2}R^{D-3}}\leq \sqrt{1-\frac{16\pi G_{D}M}{(D-2)c^2\Omega _{D-2}R^{D-3}}-%
\frac{2\Lambda _{D}R^{2}}{(D-1)(D-2)}}+1.
\end{equation}%
\end{widetext}
Hence, the inequality (%
\ref{ineqdef}) yields the upper and lower bounds
\begin{equation}
(D-2)u_{-}\leq \frac{8\pi G_{D}M}{c^2\Omega _{D-2}R^{D-3}}\leq (D-2)u_{+},
\label{masbde}
\end{equation}%
where
\begin{widetext}
\begin{equation*}
u_{\pm }=\frac{D-2}{(D-1)^{2}}\left[ 1-(1+w)\frac{D-1}{(D-2)^{2}}\Lambda
_{D}R^{2}\right] \pm \frac{D-2}{(D-1)^{2}}\sqrt{1+2w\frac{D-1}{(D-2)^{3}}%
\Lambda _{D}R^{2}}.
\end{equation*}%
\end{widetext}
For $w=-1$, where dark energy is represented by a cosmological constant, these reduce to the bounds given in Eq.~(\ref{masb}), as required. However, for $w<0$, the
upper and lower limits are more interesting, yielding different physical possibilities. First of all, both nontrivial upper and lower bounds 
only exist when
\begin{equation}
\Lambda _{D}<-\frac{1}{2wR^{2}}\frac{(D-2)^{3}}{D-1},
\end{equation}
which is the analogue of Eq. (\ref{reality_cond}). As long as the above condition is satisfied, a positive real upper bound always exists. 
Thus, any star with a mass larger than the upper bound will inevitably collapse into
a $D$-dimensional black hole.

On the other hand, for the lower bound, two possibilities exist, namely

(1) the de Sitter case, with $\Lambda _{D}>0,\ w<-(D-2)/(D-1)$.\newline

(2) the anti-de Sitter case, with $\Lambda _{D}<0,\ w>-(D-2)/(D-1)$.

For positive $\Lambda _{D}$, the accelerated expansion of the $D$%
-dimensional universe requires
\begin{equation}
w<-(D-3)/(D-1)~(=-1/3~\mathrm{for}~D=4).  \notag
\end{equation}%
The existence of a minimum mass demands the more stringent constraint, $%
w<-(D-2)/(D-1)$, which gives $w < -2/3$ for $D=4$. When the negative
pressure from the dark energy is sufficiently strong, i.e., when the
conditions in scenario (1) above are satisfied, a static star cannot be formed if
\begin{widetext}
\begin{equation*}
\frac{(D-2)^{2}}{(D-1)^{2}}\left[ 1-(1+w)\frac{D-1}{(D-2)^{2}}\Lambda
_{D}R^{2}-\sqrt{1+2w\frac{D-1}{(D-2)^{3}}\Lambda _{D}R^{2}}~\right] >\frac{%
8\pi G_{D}M}{c^2\Omega _{D-2}R^{D-3}}.
\end{equation*}
\end{widetext}

For negative $\Lambda_D$, the constraint which must be satisfied for the nontrivial minimum mass to
exist is $0>w>-(D-2)/(D-1)$. The minimum mass is given by the same condition, Eq.
(\ref{masbde}). It is interesting to note that the space-time is asymptotically
AdS in this case. The pressure from the dark energy is, however, positive
and thus helps to support the star against gravitational collapse.

For $|\Lambda_D |R^{2}\ll 1$, the generic minimum mass-radius ratio or, strictly speaking, radius
power, for both positive and negative $\Lambda_D$ becomes
\begin{equation}
\hspace{-2.6mm}\left( \frac{M}{R^{D-3}}\right) _{min}=\frac{c^2\Omega _{D-2}(-\Lambda
_{D}R^{2})}{8\pi G_{D}(D-1)}\left( 1+w~\frac{D-1}{D-2}\right) .
\end{equation}

Interestingly, the minimum mass-radius power ratio in this limit can be cast in terms of the minimum density in any number of dimensions as
\begin{eqnarray}
\hspace{-4mm}\rho_{min}& = & \frac{(D-1)M}{\Omega_{D-2}R^{D-1}}\Big\vert_{min}
 =  -\frac{c^2 \Lambda_{D}}{8\pi G_{D}}\left( 1+w\frac{D-1}{D-2}\right).\nonumber\\
\end{eqnarray}

On the other hand, the maximum mass-radius power ratio for $|\Lambda_D| R^{2}\ll 1$ is given by
\begin{eqnarray}
\left( \frac{M}{R^{D-3}}\right) _{max}&=&\frac{c^2\Omega _{D-2}}{8\pi G_{D}}\frac{(D-2)^{2}}{(D-1)^{2}}\Bigg[ 2-\nonumber\\
&&\frac{D-1}{D-2}\Lambda_{D}R^{2}\left(1+w~\frac{D-3}{D-2}\right)\Bigg]. \nonumber
\end{eqnarray}
This is greater than the Buchdahl limit for $\Lambda_{D} < 0$ and $w>-(D-2)/(D-3)$, and vice versa.

\section{Jeans instability of the dark energy fluid in arbitrary space-time dimensions} \label{Jeans}

An interesting physical possibility is that the dark energy fluid, satisfying an equation of state $p_{DE}=w\rho _{De}$ with $w=w_0=-1$, could \emph{condense gravitationally} to form stellar type stable compact objects, satisfying the same equation of state as the initial medium, but with a different parameter $w\neq -1$. The condensation process can also be described phenomenologically as the result of a viscous type dissipation process, which triggers the transition  between the two fluids.  Therefore, to study the dark energy condensation process in
the linear Newtonian regime we need to include  dissipative effects into the fluid dynamical description of the transition. Hence, we assume that the dark energy fluid has an initial density $\rho _{DE}^{(0)}=\Lambda _Dc^{2}/8\pi G_D$
and pressure $p_{DE}^{(0)}$, which satisfy the equation of state $\rho
_{DE}^{(0)}c^{2}+p_{DE}^{(0)}=0$. The fluid also has dissipative properties, characterized by the bulk
viscosity $\xi _0$, and the shear or first viscosity $\eta _0$ \cite{LaFM}, describing the internal `friction' of the decaying dark energy. The possibility that dark energy may have some anisotropic stresses, which can be  modeled with the help of a viscosity parameter, in addition to the standard sound speed equation of state parameters was considered in \cite{visc0}.

We assume that the dark energy condenses, or experiences a phase transition into a non-relativistic
  fluid, which can be characterized by a density
$\rho _{DE}$, a pressure $p_{DE}=w\rho _{DE}$, a velocity $\vec{v}$, a gravitational
acceleration $\vec{g}$, a bulk viscosity coefficient $\xi $, and a shear viscosity coefficient $\eta $.

\subsection{Hydrodynamical and first order perturbation equations}

In the following analysis, we adopt the Newtonian approximation, in which the dynamical evolution of the dark energy fluid is described by
the continuity equation, the hydrodynamical Navier-Stokes equation, and the
Poisson equation. For spherically symmetric systems in $D$ space-time dimensions, with $\left(t,r\right)\in \Re \times \Re ^{D-1}$,  the equations of motion for the dissipative dark energy fluid can therefore be written as
\begin{equation}\label{cont1}
\frac{\partial \rho }{\partial t}+\frac{1}{r^{D-2}}\frac{\partial }{\partial r}\left(r^{D-2}\rho v\right)=\frac{\partial \rho }{\partial t}+\nabla \cdot \left( \rho \vec{v}\right) =0,
\end{equation}
\bea\label{hydr1}
&&\frac{\partial v}{\partial t}+v\frac{\partial v}{\partial r}=\frac{\partial \vec{v}}{\partial t}+\left( \vec{v}\cdot \nabla \right) \vec{v}=-%
\frac{1}{\rho _{DE}}\nabla p_{DE}+\vec{g}+\nonumber\\
&&\frac{\eta }{\rho }\Delta \vec{v}+\frac{1}{\rho}\left(\xi +\frac{\eta }{3}\right)\nabla \left(\nabla\cdot \vec{v}\right),
\eea
and
\begin{equation}\label{hydr2}
\frac{1}{r^{D-2}}\frac{\partial }{\partial r}\left(r^{D-2}g\right)=\nabla \cdot \vec{g}=-8\pi\left(\frac{D-3}{D-2}\right) G_D\rho _{DE}.
\end{equation}

The coefficient on the right-hand side of Eq.~(\ref{hydr2}), derived from Gauss' law, is chosen so that this is consistent with the Einstein field equations in $D$ dimensions, Eqs.~(\ref{efeqn}).  The gravitational acceleration satisfies the condition $\nabla \times \vec{g}=0$. As the initial (unperturbed) phase of the dark energy fluid, we consider the phase characterized by the absence of the  gravitational forces, $\vec{g}=%
\vec{g}_{0}=0$, and the absence of hydrodynamical flow, which implies
$\vec{v}=\vec{v}_{0}=0$.  Moreover, we assume
constant values for the density and pressure, so that $\rho _{DE}=\rho _{DE}^{(0)}$ and $p_{DE}=p_{DE}^{(0)}$.
The dark energy phase transition, or condensation
process, leads to the generation of a gravitational interaction in the dark energy fluid,  and to small perturbations of the
hydrodynamical and thermodynamical quantities, so that
\begin{eqnarray*}
\rho _{DE}&=&\rho _{DE}^{(0)}+\rho _{DE}^{(1)}, \ \ p_{DE}=p_{DE}^{(0)}+p_{DE}^{(1)},\ \ \vec{v}=\vec{v}_{0}+\vec{v}_{1}, \nonumber\\
 \vec{g}&=&\vec{g}_{0}+\vec{g}_{1}, \;\;\;\;\;\;\;\;\;\;\;\;\;\;\;\xi =\xi _0+\xi _1, \;\;\;\;\;\;\;\eta =\eta _0+\eta _1,
\end{eqnarray*}
with the perturbed quantities satisfying the conditions $-1\ll \rho _{DE}^{(1)}/\rho _{DE}^{(0)}\ll 1$, $-1\ll p_{DE}^{(1)}/p_{DE}^{(0)}\ll 1$, $\xi _1 \ll \xi _0$, and $\eta _1 \ll \eta _0$, respectively.
In the first order of approximation, 
Eqs. (\ref{cont1}) and (\ref{hydr1}) take the form
\begin{equation} \label{hydr3}
\frac{\partial \rho _{DE}^{(1)}}{\partial t}+\rho _{DE}^{(0)}\nabla \cdot
\vec{v}_{1}
=0,
\end{equation}
\bea\label{hydr3_n}
\frac{\partial \vec{v}_{1}}{\partial t}&=&-\frac{v_{s}^{2}}{\rho _{DE}^{(0)}}%
\nabla \rho _{DE}^{(1)}+\vec{g}_{1}+\frac{\eta _0}{\rho _{DE}^{(0)}}\nabla ^2 \vec{v}_1+\nonumber\\
&&\frac{1}{\rho _{DE}^{(0)}}\left(\xi _0+\frac{\eta _0}{3}\right)\nabla \left(\nabla \cdot \vec{v}_1\right),
\eea
with
\begin{equation}
\nabla \times \vec{g}_{1}=0,\ \ \ \nabla \cdot \vec{g}_{1}=-8\pi\left(\frac{D-3}{D-2}\right) G_D\rho
_{DE}^{(1)},
\end{equation}
where we have introduced the adiabatic sound speed $v_{s}$ in
the condensed dark energy fluid, defined as $v_{s}=\sqrt{p_{DE}^{(1)}/\rho _{DE}^{(1)}}=\sqrt{%
\partial p_{DE}^{(1)}/\partial \rho _{DE}^{(1)}}=\sqrt{w}$, with $w\neq w_0$. It is important to note that the requirement that the speed of sound in the dark energy condensate be real imposes the condition $w>0$ for the newly formed fluid. This implies the physically important result that {\it the dark energy can condense only into a `normal' matter fluid}, satisfying a linear barotropic equation of state.

Next, by taking the partial derivative of the continuity equation, Eq. (\ref{hydr3}), with respect
to the time, and with the use of Eq.~(\ref{hydr3_n}), we
obtain the equation describing the propagation of the density perturbations in the condensed dark energy fluid as
\bea\label{hydrf0}
\frac{\partial ^{2}\rho _{DE}^{(1)}}{\partial t^{2}}&=&v_{s}^{2}\nabla ^{2}\rho _{DE}^{(1)}+%
\frac{(D-3)\Lambda _Dc^{2}}{D-2}\rho _{DE}^{(1)}+\nonumber\\
&&\Theta _0\frac{\partial}{\partial t}\nabla ^2 \rho _{DE}^{(1)}+\left(\Xi _0+\frac{\Theta _0}{3}\right)\frac{\partial}{\partial t}\nabla ^2 \rho _{DE}^{(1)},\nonumber\\
\eea
where we have introduced the notations
\be
\Theta _0:=\frac{\eta _0}{\rho _{DE}^{(0)}},\Xi_0:=\frac{\xi _0}{\rho _{DE}^{(0)}}.
\ee

\subsection{Condensation of the ideal dark energy fluid}

If we neglect the viscous type dissipative effects in the dark energy fluid, that is, if we assume that $\Xi _0=\Theta _0=0$, the equation describing the propagation of the density perturbations takes the form
\be\label{hydrf}
\frac{\partial ^{2}\rho _{DE}^{(1)}}{\partial t^{2}}=v_{s}^{2}\nabla ^{2}\rho _{DE}^{(1)}+%
\frac{(D-3)\Lambda _Dc^{2}}{D-2}\rho _{DE}^{(1)}.
\ee

We now look for a solution of Eq.~(\ref{hydrf}) of the form $\rho _{DE}^{(1)}\propto \exp
\left[ i\left( \vec{k}_D\cdot \vec{r}-\omega _Dt\right) \right] $, where $\omega _D, \vec{k}_D={\rm constant}$.
Thus, for the angular frequency $\omega _D$, we
obtain the following dispersion relation
\begin{equation}\label{67}
\omega ^{2}_D=v_{s}^{2}\vec{k}_D^{2}-\frac{(D-3)\Lambda _Dc^{2}}{D-2}.
\end{equation}

From Eq.~(\ref{67}) we see that, for $k < k_{J}^{(D)}$, where
\begin{equation}\label{kj}
k_{J}^{(D)}=\sqrt{\frac{(D-3)\Lambda _Dc^{2}}{(D-2) v_{s}^{2}}},
\end{equation}
is the Jeans wave number,  $\omega _D$ becomes
an imaginary quantity. In this case, we have an instability in the
dark energy fluid, which implies that $\rho _{DE}^{(1)}$ can increase (or decrease) exponentially. 
This leads to either a gravitational condensation (phase transition), or to a rarefaction. Therefore, for $k_D<k_{J}^{(D)}$, it follows that  $%
\omega _D=\pm v_{s}\sqrt{k_D^{2}-k_{J}^{(D)\;2}}=i{\rm Im}\;\omega _D$, where $ {\rm Im}\;%
\omega _D=\pm v_{s}\sqrt{k_{J}^{(D)\;2}-k_D^{2}}$ and, consequently, $\rho
_{1}\propto \exp \left[ \pm \left| {\rm Im}\omega _D\right| t\right]$.

Thus, when the mass of the newly formed dark fluid phase exceeds the mass of a sphere with radius $2\pi /k_{J}^{(D)}$, a gravitational instability occurs, leading to the collapse of the system. The critical mass corresponding to the onset of the instability is the Jeans mass $M_J^{(D)}$, defined in $D$ dimensions as
\begin{widetext}
\begin{eqnarray}
M_{J}^{(D)}&=&\left( \frac{\Omega_{D-2} }{D-1}\right) \left( \frac{2\pi }{k_{J}^{(D)}}\right) ^{D-1}\rho _{DE}^{(0)}
=\frac{\Omega_{D-2}}{D-1}\frac{(2\pi)^{D-1}}{8\pi}\left( \frac{D-2}{D-3}\right)^{(D-1)/2}\left( \frac{v_{s}}{c}\right)^{D-1} \frac{c^{2}}{G_{D}}\Lambda_D^{(3-D)/2}.
\end{eqnarray}
\end{widetext}
It should be noted that, for $D=3$, the Jeans mass is independent of $\Lambda_3$.
Hence, the four-dimensional Jeans mass of the condensed phase of the dark energy fluid can be expressed as \cite{Boehm3}
\bea
M_{J}^{(4)}&=&\frac{8\sqrt{2}}{3}\pi ^{3}\left( \frac{v_{s}}{c}\right) ^{3}\frac{%
c^{2}}{G}\Lambda^{-1/2}\approx \nonumber\\
&&1.6\times 10^{30}\times \left(
\Lambda \;{\rm cm}^{-2}\right) ^{-1/2}\left(
\frac{v_{s}}{c}\right) ^{3}\;{\rm g}.
\eea
where we have written $G_4=G$, $\Lambda_4 = \Lambda$, by convention. In four dimensions, we have $\Lambda = 3\times 10^{-56}$cm$^{-2}$, and hence we obtain
\be
M_{J}^{(4)}=9.
24\times 10^{57}\left( \frac{v_{s}}{c}\right) ^{3}\; {\rm g} =4.62\times 10^{24}\times \left( \frac{v_{s}}{c}\right) ^{3}M_{\odot}.
\ee
Using the representation of the four-dimensional cosmological constant in terms
of the universal physical constants suggested in Eq.~(\ref{const}),
we obtain the equivalent expression for the critical Jeans mass of the condensed dark energy
\begin{equation} \label{M_J^(4)}
M_{J}^{(4)}=\frac{8\sqrt{2}}{3}\pi ^{3}\left( \frac{v_{s}}{c}\right) ^{3}\frac{%
e^{6}}{\hbar G^{2}m_{e}^{3}c}.
\end{equation}
The effective four-dimensional radius $R_{J}^{(4)}$ of the stable condensed dark energy system is
then given by
\bea
R_{J}^{(4)}&=&2^{3/2}\pi \frac{v_{s}}{c}\Lambda ^{-1/2}=2^{3/2}\pi \frac{v_{s}%
}{c}\frac{e^{6}}{\hbar Gm_{e}^{3}c^{3}}\approx \nonumber\\
&&5.13\times 10^{28}\times \frac{v_{s}}{c}\;{\rm cm}.
\eea

\subsubsection{Ideal Dark Energy Jeans mass as a Chandrasekhar mass}

An important result in theoretical astrophysics is the value of the maximum mass of stable
compact objects, such as white dwarfs, neutron stars and quark
stars (see \cite{quarkstars} and references therein), obtained by Chandrasekhar. This maximum mass, known as the Chandrasekhar mass $M_{Ch}$, or the Chandrasekhar limit \cite{ShTe83}, is given by the relation
\begin{equation}\label{chandra}
M_{Ch}\approx \left[ \left( \frac{\hbar c}{G}\right)
m_{B}^{-4/3}\right] ^{3/2} = \frac{m_{Pl}^3}{m_B^2},
\end{equation}
where $m_{B}$ represents the mass of the particle species giving the main
contribution to the stellar mass ~\cite {ShTe83} (for example, in the astrophysically important cases of white dwarfs
and neutron stars, $m_B$ represents the baryon mass) and $m_{Pl}$ is the reduced Planck mass. Thus, it turns out that, with the exception of
some composition-dependent numerical factors of order unity, the maximum mass of a compact star can be expressed in terms of fundamental physical constants only.

Using Eq. (\ref{M_J^(4)}), the four-dimensional Jeans mass for the condensed dark energy fluid can also be written in a form similar to the
Chandrasekhar mass if we assume that dark energy consists of particles having an
effective mass $m_{eff}$, which are given by
\begin{equation}
m_{eff} = \left(G\hbar ^{5}c^{5}\right)^{1/4}\frac{m_{e}^{3/2}}{e^{3}}.
\end{equation}
The Jeans mass of the dark energy phase is then obtained as
\bea
M_{J}^{(4)}&=&\left[ \frac{8\sqrt{2}\pi ^{3}}{3}\left( \frac{v_{s}}{c}\right) ^{3}\right] %
\left[ \left( \frac{\hbar c}{G}\right) m_{eff}^{-4/3}\right] ^{3/2} \approx \nonumber\\
&&\left( \frac{v_{s}}{c}\right)^{3}\frac{m_{Pl}^3}{m_{eff}^2}.
\eea
In this scenario, the effective mass of the elementary dark matter particles is of order of $m_{eff}\approx
8\times 10^{-20}$ g.

\subsection{The effect of the bulk viscosity on dark energy condensation}

In order to investigate the effects of the dissipative processes on the Jeans condensation of the dark energy fluid, we consider a simple case in which we neglect the shear viscosity by taking $\Theta _0\approx 0$. The only viscous dissipative effect is then given by the bulk viscosity and the equation describing the density perturbations of the dark energy fluid takes the form
\bea\label{visc}
\frac{\partial ^{2}\rho _{DE}^{(1)}}{\partial t^{2}}&=&v_{s}^{2}\nabla ^{2}\rho _{DE}^{(1)}+%
\frac{(D-3)\Lambda _Dc^{2}}{D-2}\rho _{DE}^{(1)}+\Xi _0\frac{\partial}{\partial t}\nabla ^2 \rho _{DE}^{(1)}. \nonumber\\
\eea

By looking again for plane wave solutions of Eq.~(\ref{visc}) of the form $\rho _{DE}^{(1)}\propto \exp
\left[ i\left( \vec{k}_D\cdot \vec{r}-\omega _Dt\right) \right] $, we obtain the generalized dispersion relation
\be
\omega ^{2}_D=v_{s}^{2}\vec{k}_D^{2}+i\Xi _0\vec{k}_D^{2}\omega _D-\frac{(D-3)\Lambda _Dc^{2}}{D-2}.
\ee

The $D$-dimensional dispersion relation for dark energy in the presence of bulk viscosity has the solutions
\be\label{omegaf}
\omega _D=i\frac{\Xi _0\vec{k}_D^2}{2}\pm \sqrt{\omega _D^{(0)}},
\ee
where
\be\label{79}
\omega _D^{(0)}=-\frac{\vec{k}_D^4\Xi _0^2}{4}+v_s^2\vec{k}_D^2-\frac{(D-3)\Lambda _Dc^{2}}{D-2}.
\ee

Thus, from Eq.~(\ref{omegaf}), it follows that we obtain a time dependent exponential regime for $\omega _D^{(0)}<0$, and a damped oscillatory
regime for $\omega _D^{(0)}>0$. It is important to note that the pure oscillatory Jeans condensation regime of the $D$-dimensional ideal dark energy
fluid is lost. On the other hand, the viscous effects do not change the threshold value of the Jeans mass of the dark energy fluid, though
they may drastically  modify the evolution of the perturbations.

The equation $\omega _D^{(0)}=0$ has two solutions for $k_D$, given by
\be
k _{D}^{\pm}=\frac{\sqrt{2}v_s}{\Xi _0}\left[1\pm\sqrt{1-\frac{\Xi _0^2}{v_s^4}\frac{D-3}{D-2}\Lambda _Dc^{2}}\right]^{1/2}.
\ee
and the existence of such solutions imposes the constraint
\be
\Xi _0\leq \sqrt{\frac{D-2}{(D-3)\Lambda _Dc^2}}v_s^2,
\ee
or, equivalently,
\be
\xi _0\leq \frac{1}{8\pi G_D}\sqrt{\frac{D-2}{D-3}\Lambda _Dc^2}\;v_s^2,
\ee
on the bulk viscosity coefficient of the dark energy fluid.


\section{Quantum implications of a classical minimum mass density} \label{quant}

In this Section, we consider the quantum mechanical implications of a classical minimum mass density, implied by the generalized Buchdahl inequalities in the presence of a positive cosmological constant, $\Lambda > 0$. For simplicity, we consider the $D=4$ case, in which the generalized Buchdahl inequality implies \cite{Har1,Boehmer:2005sm}
\begin{equation}\label{min_dens}
2GM \geq \frac{\Lambda c^2}{6}R^3, \ \ \ \rho = \frac{3M}{4\pi R^3} \geq \frac{\Lambda c^2}{16\pi G} =: \rho_{min},
\end{equation}
where $\rho_{min}$ denotes the minimum density.

We now define the (non-reduced) Planck length and mass scales as
\begin{equation}\label{Planck}
R_P := \sqrt{\frac{hG}{c^3}}, \ \ \ M_P := \sqrt{\frac{hc}{G}},
\end{equation}
and the (non-reduced) Wesson masses \cite{Wesson:2003qn} as
\begin{equation}\label{Wesson}
M_W := \frac{h}{c}\sqrt{\frac{\Lambda}{3}} = M_P\frac{R_P}{R_{W}}, \ \ \ M'_W := \frac{c^2}{G}\sqrt{\frac{3}{\Lambda}}
= \frac{M_P^2}{M_W},
\end{equation}
where we have parameterized $\Lambda$ in terms of a characteristic length scale via
\begin{equation}\label{Lambda}
\Lambda =: \frac{3}{R_{W}^2}.
\end{equation}
In his original definitions \cite{Wesson:2003qn}, Wesson used $\hbar$ instead of $h$ but, for convenience, we adopt the definitions in Eq. (\ref{Wesson}) and use the non-reduced Planck scales, since these are analogous to the standard definition of the Compton wavelength using $h$, not $\hbar$. Hence, $R_W$ is just the Compton wavelength associated with the first Wesson mass, $M_W$. The approximate value of the second Wesson mass is therefore $M'_W \approx 1.347 \times 10^{56}$ g, which may be interpreted as the mass of the observable universe, whereas it has been suggested that $M_W \approx 1.809 \times 10^{-65}$ g should be interpreted as a minimum mass scale in nature \cite{Har1,Boehmer:2005sm,Wesson:2003qn}.

Another way to derive $M'_{W}$ is to note that the temperature and entropy of the de Sitter cosmic horizon are given by~\cite{Shankaranarayanan:2003ya}
\be
k_{B}T=\frac{3(\hbar c)}{2\pi R_{W}},
\ee
 and
 \be
 S=\frac{4\pi R^{2}_{W}k_{B}c^{3}}{12G\hbar},
 \ee
respectively.  Using the Smarr formula for a non-rotating uncharged black hole,
\begin{equation}
M _{BH}= \frac{2TS}{c^{2}}, \nonumber
\end{equation}
we obtain the mass of the corresponding black hole as
\begin{equation}
M_{BH} = \frac{c^{2}}{G}\sqrt{\frac{3}{\Lambda}}, \nonumber
\end{equation}
which is simply the second (non-reduced) Wesson mass $M'_{W}$.  Thus, this mass can be naturally interpreted as the maximum possible mass contained within the de Sitter horizon, above which the observable universe would collapse to form a cosmic black hole. In addition, as $R_W \approx 10^{26}$ cm is of the order of the current horizon radius, we may interpret the proposed bound $M \geq M_W $ in another way, at least at the present epoch, as stating that no particle can have a Compton wavelength larger than the radius of the universe.

However, combining the classical bound on the mass density, Eq. (\ref{min_dens}), with the Compton bound on the radius of a quantum mechanical object
\begin{equation}\label{Compton}
R \geq R_C := \frac{h}{Mc},
\end{equation}
leads to a new prediction of a minimum mass scale:
\begin{equation}\label{M_Lambda}
M \geq M_{\Lambda} := \frac{1}{\sqrt{2}}\sqrt{M_PM_W}.
\end{equation}
This implies the existence of a maximum Compton radius:
\begin{equation}\label{R_Lambda}
R_C \leq R_{\Lambda} := \sqrt{2}\sqrt{R_PR_W}.
\end{equation}

These new critical scales, $M_{\Lambda}$ and $R_{\Lambda}$, are functions of $M_P$ and $M_W$ or $R_P$ and $R_W$, respectively. That is, unlike the Wesson masses defined in Eq. (\ref{Wesson}), or their associated length scales, and unlike the Planck scales Eq. (\ref{Planck}), each depends on all four `universal' constants $G$, $c$, $h$ and $\Lambda$, simultaneously. Using the approximate values given above, the new scales can be estimated as
\begin{equation}\label{estimates}
R_{\Lambda} \approx 9.002 \times 10^{-3}\; {\rm cm}, \ \ \ M_{\Lambda} \approx 2.456 \times 10^{-35}\; {\rm g}.
\end{equation}
For comparison, the current upper bound on the average neutrino mass obtained from the Planck mission data is $M_{\nu} \leq 0.23$  eV $= 1.8 \times 10^{-33}$ g \cite{Planckresults}

We may also associate a temperature with a given mass scale via the identification $T \sim Mc^2/k_B$, where $k_B \approx 1.381 \times 10^{-23} {\rm JK^{-1}}$ is Boltzmann's constant. Interestingly, defining the temperature $T_{\Lambda}$ associated with the minimum mass $M_{\Lambda}$ via
\begin{equation}\label{T_Lambda}
T_{\Lambda} := \frac{1}{16\pi}\frac{M_{\Lambda}c^2}{k_B},
\end{equation}
gives $T_{\Lambda}  \approx  3.18 {\rm K}$. This is remarkably close to the current temperature of the CMB radiation, though the numerical value of the constant of proportionality is somewhat arbitrary, here being chosen to match the numerical factor in the Einstein-Hilbert action.

Nonetheless, it is interesting that this mass scale has previously been proposed as a minimum mass for stable dark matter relics, based on loop quantum gravity calculations \cite{Carr:2015nqa}. Sub-Planck mass loop black hole (LBH) solutions have been shown to exist if quantum gravity effects give rise to a quadratic generalized uncertainty principle (GUP) \cite{Carr:2011pr}. In this case, the usual relations between the black hole mass, its horizon and its temperature invert at $M \sim M_P$, so that $R_S \propto M^{-1}$ and $T_{H} \propto M$ for $M < M_P$. Such objects behave like `black atoms', but continue to decay via radiation emission until they reach thermal equilibrium with the CMB photon bath \cite{Carr:2011pr} (see also \cite{Nicolini:2008aj} and references therein).

We can also use the fact that the Planck density must be greater than the minimum classical density, $\rho_{P} \geq \rho_{min}$, to place theoretical constraints on the values of $M_W$ and $R_W$ relative to $M_P$ and $R_P$, respectively, and hence on the value of $\Lambda$ in terms of the other three constants $G$, $c$ and $h$. The Planck density is defined as
\begin{equation}\label{Planck_dens}
\rho_P := \frac{3M_P}{4\pi R_P^3} = \frac{3}{4\pi}\frac{c^5}{hG^{2}},
\end{equation}
so that
\begin{equation}\label{W-bounds1}
R_W \geq \frac{1}{2}R_P, \ \ \ M_W \leq 2M_P.
\end{equation}
These conditions are satisfied (by many orders of magnitude) for the actual values of $R_W$, $R_P$, $M_W$ and $M_P$ but it is interesting to note that, regardless of the measured values of $G$, $c$, $h$ and $\Lambda$, they leave two possible scenarios for stable compact objects which behave quantum mechanically:
\begin{eqnarray}\label{W-bounds2}
R_P \leq R_C \leq R_{\Lambda} \iff M_P \geq M \geq M_{\Lambda},
\end{eqnarray}
\begin{eqnarray}\label{W-bounds3}
R_C \leq R_P \leq R_{\Lambda} \iff M \geq M_P \geq M_{\Lambda}.
\end{eqnarray}
The horizons of a Schwarzschild-de Sitter black hole are defined by the equation
\begin{eqnarray}\label{Schwarz-dS_1}
R^3 - R_W^2R + R_W^2R_S \geq 0,
\end{eqnarray}
so that the inner horizon is of the order of the standard Schwarzschild radius $R_S$, whereas the outer horizon is comparable to the horizon of the universe, $R_W$. Therefore, Eq. (\ref{W-bounds2}) corresponds to the realm of elementary particles, whereas Eq. (\ref{W-bounds3}) corresponds to objects with the potential to form black holes, if compressed beyond the critical radius implied by Eq. (\ref{Buch_1}).

Eq. (\ref{W-bounds2}) may have profound implications, since it predicts the existence of a phenomenologically significant length scale $R_{\Lambda}$ which, we may conjecture, demarcates the boundary between quantum mechanical and classical behavior. This may have implications for the study of gravitational decoherence. Surprisingly, despite the common interpretation of the cosmological constant as a `gravitational' phenomenon, relatively little work had been done on this topic in the context of models with $\Lambda \neq 0$ (see \cite{GravDec} and references therein).

Furthermore, setting
\begin{eqnarray}\label{}
R_W \geq R_{min} := \frac{9}{4}\frac{R_PM}{M_P}
\end{eqnarray}
implies
\begin{eqnarray}\label{}
M \lesssim M'_W := \frac{M_P^2}{M_W},
\end{eqnarray}
which bolsters the claim that $M'_W$ should be interpreted as a maximum mass.  By contrast, the classical minimum radius for a gravitationally stable object may either be above or below the maximum Compton wavelength: $R_{min} \leq R_{\Lambda}$ implies that the system behaves like a quantum mechanical `particle', whereas $R_{min} \geq R_{\Lambda}$ corresponds to the classical regime. Using Eq. (\ref{Buch_1}) and defining
\begin{eqnarray}\label{}
M'_{\Lambda} := M_P^2/M_{\Lambda},
\end{eqnarray}
the different regimes may also be defined via the mass of the system, so that
\begin{eqnarray}\label{}
R_{min} \geq R_{\Lambda} \iff M \gtrsim M'_{\Lambda} 
\nonumber\\
R_{min} \leq R_{\Lambda} \iff M \lesssim M'_{\Lambda}. 
\end{eqnarray}

In summary, the results obtained above suggest that  systems with masses in the range:
\begin{itemize}

\item $M_{\Lambda} \leq M \lesssim M_P$ correspond to elementary particles, whose behavior is manifestly quantum mechanical

\item $M_P \lesssim M \lesssim M'_{\Lambda}$ behave quantum mechanically but have the potential to form black holes (which continue to behave like quantum `particles') if compressed below their Buchdahl limit

\item $M'_{\Lambda} \lesssim M \lesssim M'_W$ behave classically and have the potential to form black holes (which continue to behave like classical particles) if compressed below their Buchdahl radius.

\end{itemize}
In this model, both fundamental particles with mass $M_{\Lambda} \leq M \lesssim M_P$ and stable compact objects with masses $M_P \lesssim M \lesssim M'_{\Lambda}$ behave quantum mechanically: what they have in common is that both have radii $R \leq R_{\Lambda}$. These conclusions follow from a simple combination of results from classical general relativity and ordinary quantum mechanics.

We can also interpret the new length and mass scales in the following way. Let us begin by defining the minimum classical energy density, given in Eq. (\ref{min_dens}), as the energy density associated with the cosmological constant which, in turn, is the minimum possible energy density of space-time, $\rho_{\Lambda} := \rho_{min}$. Next, we interpret this as a `sea' of dark energy particles, each with effective mass $M_{\Lambda}$ and associated Compton wavelength $R_{\Lambda} = R_PM_P/M_{\Lambda}$. Furthermore, we assume that space is saturated with such particles and that the de Sitter vacuum expands in such a way as to maintain the associated energy density, so that
\begin{equation}\label{}
\rho_{\Lambda} := \frac{3\Lambda c^2}{4\pi G} = \frac{3M_{\Lambda}}{4\pi R_{\Lambda}^3}.
\end{equation}
Substituting for $M_{\Lambda}$ and $R_{\Lambda}$, respectively, then yields the order of magnitude values defined in  Eqs.~(\ref{R_Lambda})-(\ref{M_Lambda}).

Hence, by combining standard Compton type arguments, which imply a minimum radius for a quantum object with a given mass, with classical Buchdahl type bounds for $\Lambda >0$, which imply a minimum mass density, we are led naturally to a picture in which dark energy is composed of a sea of quantum particles with effective mass $M_{\Lambda} \propto \Lambda^{1/4}$. The associated mass density $\rho_{\Lambda}$ is given by $M_{\Lambda}$ divided by the volume occupied by each particle due to its Compton wavelength. In this picture, the dark energy condensation picture discussed in Sec. \ref{Jeans} occurs due to fluctuations which lower the effective mass within a localized region, leading to local over-densities. This is equivalent to a local softening of the effective equation of state.

\begin{figure}[h]
 \centering
       \includegraphics[width=0.45\textwidth]{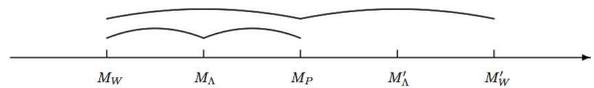}
       \caption{ A schematic representation of the hierarchy of mass scales, related by geometric means.  } \label{fig1}
\end{figure}

It is interesting to observe that in the hierarchy of mass scales $M_{W} < M_{\Lambda} < M_{P} < M'_{\Lambda} < M'_{W}$, each mass is all related to its `neighbors' by the geometric mean $M_{i}\simeq \sqrt{M_{i-1}M_{i+1}}$, for $M_{i-1}<M_{i}<M_{i+1}, \  i =1,2,..,5$, where $\{M_{1},M_{2},M_{3},M_{4},M_{5}\} :=\{M_{W},M_{\Lambda},M_{P},M'_{\Lambda},M'_{W}\}$.  An illustration~(not to scale) of the mass scale hierarchy is given in Fig.~\ref{fig1}.
Since the geometric relations are {\it transitive} among $M_{i-1},M_{i},M_{i+1}$ and $M_{i-2},M_{i},M_{i+2}$ triplets, Fig.~\ref{fig1} also implies that $M'_{W} \simeq M'^{2}_{\Lambda}/M_{P}$ which is straightforward to verifiy directly. We note that these geometric relations originate from the following three physical constraints: (1) the size of a classical object is larger than or equal to its Compton wavelength, (2) both a minimum mass and minimum density exist, in general relativity, in space-times with a positive cosmological constant, $\Lambda > 0$, and (3) the maximum mass of a gravitationally stable classical compact object also exists in this scenario.

Remarkably, these mass scales also admit other physical interpretations. As stated above, there exist both sound theoretical reasons and empirical evidence to support the claim that $M'_W$ should be interpreted as the maximum possible mass of a de Sitter universe. In addition, it is straightforward to verify that the Chandrasekhar mass for a condensate of particles of mass $M_{\Lambda}$ is also equal, up to numerical coefficients of order unity, to $M'_W$. Since we require $M \lesssim M'_W$, this is is another way of saying that our (known) universe cannot collapse and must expand forever, as implied by observations suggesting a phase of accelerated expansion beginning at the current epoch \cite{Planckresults,Bet}.

Furthermore, if we assume that dark matter particles, with mass $M_{\Lambda}$, formed from the condensation of dark energy, this implies that the \emph{initial} dark matter temperature, at the epoch of formation, is equal to the (constant) temperature of the dark energy fluid in its original phase. Even if this is not the case, by envisaging dark energy as a `sea' of quantum mechanical particles, the onset of accelerated expansion then coincides with the point at which the temperature of the thermal bath of photons drops below the temperature associated with the effective mass of the dark energy particles, $M_{\Lambda}$, for the first time. It would be interesting to further investigate this in the context of thermodynamic interpretations of gravity (see, for example, \cite{Davies:1980hz,Padmanabhan:2009vy,Clifton:2013dha} and references therein). In particular, one may hope to obtain a `thermodynamic interpretation' of the coincidence problem, posed by the onset of accelerated expansion at the present time.

We also note that the expression for the minimum radius, $R \geq R_{min} = (R_P^2R_W)^{1/3} \approx 10^{-15}$ m, which is of the same order of magnitude as the classical electron radius $r_e$, and which was obtained previously, by two different methods, in \cite{Boehm3,Beck:2008rd}, may also be obtained a third way by requiring the density of the Universe to be less than the Planck density, $(3/4\pi)M'_W/R^3 \leq (3/4\pi)M_P/R_P^3 =: \rho_P$. In this context, $R_{min} = (R_P^2R_W)^{1/3} \approx r_e$ may be interpreted as the minimum classical radius to which the known universe may be compressed before it exceeds the Planck density. 

Finally, before concluding this Section, we note that, using Eq.~(\ref{masb}),  we may obtain a further generalization of Eq.~(\ref{min_dens}) which is valid in arbitrary dimensions. An analysis similar to that given above should then yield results valid for any value of $D \geq 2$. However, in this case a subtlety may arise, since it has recently been proposed in \cite{LC1} that, for space-times with compact dimensions, the Compton wavelength changes on scales $R_C < R_E$, where $R_E$ is the length scale of the compactification. We therefore leave a full analysis of the higher-dimensional case, including both spherically symmetric and non-spherically symmetric space-times, to a later publication.

\section{Conclusions and Discussions}\label{concl}

The existence of dark energy, as proved by a plethora of astrophysical and cosmological observations \cite{Planckresults,Bet}, has fundamentally modified the landscape of theoretical physics. If dark energy, represented by a cosmological constant, is one of the major components of the Universe, it is a reasonable assumption to include it among the fundamental constants of nature \cite{Wesson:2003qn}. Hence, we can extend the set of fundamental constants, which can be taken as the speed of light $c$, the gravitational constant $G$, Planck's constant $h$, and the cosmological constant $\Lambda$. Therefore, the mere existence of the cosmological constant, or, at least, of some form of dark energy, may imply drastic modifications or extensions of the basic laws of physics.

If the set of fundamental constants is enlarged, it follows that there are two different masses that can be constructed from $c$, $G$, $h$ and $\Lambda$ \cite{Wesson:2003qn}. The first Wesson mass, $M_W=\left(h /c\right)\sqrt{\Lambda /3}\approx 10^{-66}$ g, may be relevant at the quantum scale, while the second Wesson mass, $M'_W=\left(c^2/G\right)\sqrt{3/\Lambda}\approx 10^{56}$ g, has the same order of magnitude value  as the observable mass of the Universe.

From a theoretical point of view, $M'_W$ can be obtained as the upper bound on the mass of a gravitationally stable spherically symmetric object in four-dimensional general relativity in the presence of a positive cosmological constant, $\Lambda > 0$, and as the Jeans mass of a gravitationally unstable dark energy condensate, respectively. Alternatively, the Smarr formula for uncharged non-rotating black holes suggests that the mass of the Universe, $M_U \approx M'_{W}$, follows from thermodynamic properties of the de Sitter horizon.

In the present paper, we have extended previous results concerning the Buchdahl, Jeans and minimum mass limits, derived initially in four space-time dimensions with $\Lambda > 0$, to arbitrary $D$-dimensional, static, spherically symmetric geometries with generic $\Lambda_D \neq 0$. We have shown that such limits exist and are well defined for $D \geq 4$. Moreover, the existence of a minimum mass is closely related to the existence of a minimum density. From the investigation of the Jeans instability in the framework of a simple Newtonian model, we obtained $M'_W$ as the critical Jeans mass of the dark energy fluid, corresponding to the mass of gravitationally stable dark energy objects. Using this result, we were also obtain a new physical interpretation of the cosmological constant in arbitrary space-time dimensions. From Eq.~(\ref{kj}), by taking the speed of sound $v_s$ in the gravitationally condensed dark energy fluid as approximately equal to the speed of light, $v_s \approx  c$, we obtain for the $D$-dimensional cosmological constant the expression $\Lambda _D \approx \left[(D-2)/(D-3)\right]\left[k_J^{(D)}\right]^2$, where $k_J^{(D)} = 2\pi/R_J^(D)$ is the $D$-dimensional Jeans radius. Hence, from a  physical point of view, the $D$-dimensional cosmological constant represents the square of the $D$-dimensional Jeans wave number of the dark energy fluid embedded in a higher dimensional geometry.
In addition, with the use of Eq.~(\ref{const}), the mass of the Universe $M_{U}$ can be expressed, in four dimensions, in terms of the fundamental constants $\left(c,G,\hbar,e,m_e\right)$ as
\be
M'_{W}\approx M_{U}\approx \frac{\sqrt{3}e^6}{cG^2\hbar m_e^3} \approx\frac{r_e^3 m_{Pl}}{l_{Pl}^3}.
\ee

where $l_{Pl}=\sqrt{\hbar G/c^3}$ and $m_{Pl}=\sqrt{\hbar c/G}$ are the reduced Planck length and mass scales and $r_e$ and $m_e$ are the classical electron radius and rest mass, respectively.

In addition, we investigated the quantum mechanical implications of the existence of both a classical minimum mass and a minimum density. By combining simple Compton type arguments for the minimum radius of a quantum mechanical object, $R_C$, with the classical minimum radius, $R_{min}$ (and assuming  $R_C \geq R_{min}$) we obtained a new minimum mass scale, $M_{\Lambda} \sim \sqrt{M_PM_W} \sim 10^{-35}$ g, where $M_P$ denotes the (non-reduced) Planck mass, which depends on all four `universal' constants, $G$, $c$, $h$ and $\Lambda$, simultaneously. Interestingly, the temperature associated with this mass scale is of the order of the present day CMB temperature. The associated length scale $R_{\Lambda} \sim \sqrt{R_PR_W} \sim 10^{-3}$ cm, where $R_P$ denotes the (non-reduced) Planck length, represents the maximum possible Compton wavelength of a quantum mechanical object, suggesting an absolute maximum decoherence length associated with `gravitational' decoherence through the interaction of the system with omnipresent dark energy.

Furthermore, these considerations lead naturally to a model in which dark energy, represented at the classical level by a cosmological constant, exists as a `sea' of quantum mechanical particles, each with effective mass $M_{\Lambda}$ and associated Compton wavelength $R_{\Lambda}$. In this scenario, the de Sitter vacuum expands so as to keep the energy density of the particle `sea', $\rho_{\Lambda} := (3/4\pi)(\Lambda c^2/G) = (3/4\pi)(M_{\Lambda}/R_{\Lambda}^3)$, constant, and the dark energy condensation process corresponds to local fluctuations in the effective mass, which is equivalent to local softening of the equation of state. Interestingly, in the hierarchical set of mass scales thus obtained, $\{M_{1},M_{2},M_{3},M_{4},M_{5}\} :=\{M_{W},M_{\Lambda},M_{P},M'_{\Lambda},M'_{W}\}$, each mass is simply the geometric mean of its nearest neighbors,  $M_{i} = \sqrt{M_{i-1}M_{i+1}}$, for $i =1,2,..,5$, providing an elegant, though as yet unexplained, geometric relation between the series of fundamental, phenomenologically significant mass scales in nature.  Remarkably, by using dimensional analysis, the geometric relations can be shown to originate from the unique dimensionless quantity
\begin{equation}
\frac{M_{2}}{M_{1}} \simeq \frac{M_{3}}{M_{2}} \simeq \frac{M_{4}}{M_{3}} \simeq \frac{M_{5}}{M_{4}} \simeq \left( \frac{c^{3}}{\hbar G\Lambda}\right)^{1/4},  \label{dimrat}
\end{equation}
constructed from the four `fundamental' constants.  Namely, neighboring masses are separated  {\it universally} by roughly $10^{30}$ orders of magnitude.  Therefore, we propose that
\be
\left( \frac{\hbar G\Lambda}{c^{3}}\right)\simeq R_P^2\Lambda \sim \frac{R_P^2}{R_W^2} \sim \frac{M_W^2}{M_P^2} \sim \frac{M_W}{M'_W} \approx 10^{-120},
\ee
should be taken as the fundamental dimensionless constant. It is the order of magnitude discrepancy between the expected value of the vacuum energy (assuming a quantum gravity theory with an energy scale characterized by the Planck mass) and the observed value.
It also has an interpretation as $N^{-1}$, where $N$ is the number of fundamental mass quanta in the universe. In other words, if we assume that all particles are made from fundamental mass quanta of the order of the first Wesson mass, $M_W$, there must be $N \sim 10^{120}$ of these particles in the current horizon radius.

In this case, there are
\be
n_{\Lambda} = \frac{M_{\Lambda}}{M_W} = \sqrt{\frac{M_P}{M_W}} = N^{1/4} \sim  10^{30},
\ee
fundamental mass quanta in each dark energy particle, and
\be
N_{\Lambda} = \left(\frac{R_W}{R_{\Lambda}}\right)^3 = \left(\frac{M_P}{M_W}\right)^{3/2} = \frac{M'_W}{M_{\Lambda}} = N^{3/4} \sim 10^{90},
\ee
dark energy particles within the horizon at the present epoch. Hence, the above considerations may `explain' (or at least interpret from a simple physical point of view) the multiplying factors between the different mass scales in the particle mass hierarchy.

Another interpretation of the dimensionless quantity $R_{P}^2 \Lambda$ is the ratio between Planck area and the area of the cosmic horizon. Holographically, this is the number of quantum gravity bits present on the boundary. The fact that the total number of bits on the boundary is equal to the total number of quanta in the bulk space indicates that holography is at work in the entire universe.

Finally, we note that the existence of mass/density bounds in the asymptotically AdS case, when the dark energy satisfies the conditions $w>-(D-2)/(D-1), \ \Lambda_{D}<0$, has interesting implications from the viewpoint of holographic duality.  The maximum mass bound for a given radius guarantees that any object with larger mass will inevitably collapse to form a black hole.  Holographically, the maximum mass would correspond to the maximum temperature~(identified with Hawking temperature of the maximum mass black hole) of the dual gauge matter before the inevitable deconfinement phase transition occurs~\cite{Burikham:2012kn, Burikham:2014ova}.  On the other hand, the existence of a minimum mass and a minimum density determines the conditions under which a gravitationally stable, static object can be formed in the presence of dark energy.  If the average density of the object is too small, it will not be able to support itself gravitationally under the outward pressure.  At present, it is unclear what the gauge theory dual of this minimum mass/density should be, since the boundary space is always asymptotically AdS regardless of the mass and size of the static star located at the center.  We leave this interesting question for a future publication.

\section*{Acknowledgments}

We are grateful to the anonymous referee for comments and suggestions that helped us to significantly improve our manuscript. We would like to thank Taum Wuthicharn for pointing out relation (\ref{dimrat}).  P.B. and K.C. are supported in part by the Thailand Research Fund~(TRF), Commission on Higher Education~(CHE) and Chulalongkorn University under grant RSA5780002. M.L. is supported by a Naresuan University Research Fund individual research grant.

\appendix

\label{appa}

\end{document}